\documentclass{article}
\usepackage{graphicx} 
\usepackage{epsfig}     
\usepackage{graphicx}   
\usepackage{url}
\usepackage{natbib}     
\usepackage{amsmath}
\usepackage{subcaption}

\date{January 2025}

\begin{document}

\title{Stochastic description of UHECR interactions \\[1ex] \large Submitted for the proceedings of UHECR-2024}

\author{Leonel Morejon$^{1}$\thanks{leonel.morejon@uni-wuppertal.de}, Julian Rautenberg$^{1}$
\vspace{2mm}\\
\it $^1$Bergische Universit\"{a}t Wuppertal,\\ 
\it Gausstrasse 20, 42117 Wuppertal, Germany
}

\maketitle

\begin{abstract}
Photointeractions of ultra-high energy cosmic rays (UHECRs) in astro-physical scenarios are in general of stochastic nature and are often modeled with Monte Carlo methods to obtain the form of the distributions resulting from a sequence of interactions. These distributions are non trivial because the products resulting from each interaction as well as the number and distances covered by the secondary nuclear species are all random.
In this work, a stochastic approach based on the theory of matrix exponential distributions is employed to describe the cascade distributions analytically and illustrate their potential for tracing the individual history of UHECRs, including inside the source. This analytic description has the advantage of better precision and considerably reduced computational cost in contrast to Monte Carlo codes, while requiring the same inputs: the interaction rates, the multiplicity, and the energy distributions of secondaries from a single interaction. The description of the composition evolution from in-source to extragalactic propagation (currently performed in separate simulations in the literature) is achieved here as a continuous distribution, using a gamma-ray burst scenario inspired by the event GRB170817A. Finally, the potential for locating a source based on the reconstructed UHECR origin employing this description is discussed under simplified general assumptions.
\end{abstract}

\section{Introduction}

The interactions of cosmic rays with background photons in UHECR sources and during extragalactic propagation are fundamental for understanding the evolution of their composition before being detected at Earth. While in both cases photonuclear interactions are of the same nature, the differences in timescales, the spectrum of target photons and the presence of other energy losses have prompted different approaches to describe the interactions within sources separately from extragalactic propagation. Furthermore, the presence of both stochastic and deterministic competing interactions has been dealt with by either neglecting the stochasticity of some quantities in favor of deterministic approximations, or addressing it via complex Monte Carlo codes with inclusion of the continuous losses but at the cost of an increased computational cost.

Recently, a stochastic analytic approach has been put forward as an alternative, and applied to sources and extragalactic propagation separately \cite{Morejon:2023cpe,Morejon:2024rtq}. The probability distributions are based on the theory of Matrix Exponentials~\cite{Bladt2017} which is well suited to describe Markov processes as the sequence of nuclear disintegrations resulting from successive interactions of UHECR nuclei with surrounding photon fields. This approach allows explicitly describing the stochastic quantities with analytic expressions in both scenarios consistently, allowing a continuous description covering the transition without artificial boundaries. An advantage of this possibility is that the composition can be studied with e.g. the regular fit to UHECR spectrum and composition with injection the composition as parameters, instead of the ejected composition as is common, allowing the inclusion of source effects directly in the fit procedure. In the following, the approach is briefly outlined, and subsequently, the complete probability distribution (in-source + extra-galactic propagation) is obtained for an example of nearby source. Finally, the potential for discovery of nearby sources with very energetic UHECRs is theorized based on simplified assumptions.

\section{Description of the approach}

Ignoring the effect of continuous energy losses (e.g. synchrotron losses, pair production losses, etc.), UHECR photonuclear interactions, such as photodisintegration and photomeson, preserve the boost of the parent nucleus and the nuclear products (with mass number $A > 1$). The resulting cascades can be described as a continuous-time Markov chain (CTMC)~\cite{Bladt2017} which corresponds to a Markov process with intensities given by the interaction rates 
\begin{equation}
    \lambda(\gamma, z) = \frac{1}{2\gamma^2}\int_0^{\infty} \frac{n(\epsilon, z)}{\epsilon^2} d\epsilon \int_0^{2\epsilon \gamma} \varepsilon \sigma (\varepsilon) d\varepsilon
    \label{eq:interaction_rate}
\end{equation}
where $\gamma$ is the Lorentz boost of the interacting cosmic ray, the photon number density of the target background photons is $n(\epsilon, z)$ (it can also depend on the redshift $z$) and the photonuclear cross sections $\sigma (\varepsilon)$ corresponds to the production of a given species via photodisintegration or photomeson interactions. In this process, each state corresponds to a nuclear species and the evolution of the occupation probabilities $\mathbf{p}(t) = \mathbf{p}(0) e^{\mathbf{\Lambda}t}$ where $\mathbf{p}(0)$ is the vector of initial occupation probabilities and the interaction matrix
\begin{equation}
    \mathbf{\Lambda}(\gamma) = 
    \begin{pmatrix}
        -\lambda_{S_1} & \lambda_{S_1 \to S_2} & \lambda_{S_1 \to S_3} & \lambda_{S_1 \to S_4}& ... & \lambda_{S_1 \to S_N}\\
        0 & -\lambda_{S_2} & \lambda_{S_2 \to S_3} & \lambda_{S_2 \to S_4} & ... & \lambda_{S_2 \to S_N}\\
        0 & 0 & -\lambda_{S_3} & \lambda_{S_3 \to S_4} & ... & \lambda_{S_3 \to S_N}\\
        ... & ... & ... & ... & ... & ... &\\
        0& 0& 0& 0& ... &-\lambda_{S_N}\\        
    \end{pmatrix}
\end{equation}
contains all the rates for production of a certain species by any other evaluated at $\gamma$.

The presence of continuous energy losses produces changes in the boost, and therefore implies that the rates vary as the cascade develops. This corresponds to the more general inhomogeneous continuous-time Markov chains (ICTMCs) which do not have a general analytic expression. However, some changes of the interaction rates can be expressed analytically, such as the case where the boost evolves independently of the sequence of the cascade (or the occupation probability vector). In these situations the interaction matrix can be written as $\mathbf{\Lambda}(\gamma, t) = \mu(t)\mathbf{\Lambda}(\gamma)$ and the evolution of the occupation probabilities preserves its analytic form with different dependence: $\mathbf{p}(t) = \mathbf{p}(0) e^{\mathbf{\Lambda} \int_0^t \mu(s) ds }$ . An example of this type of situation is the case where the target photon field is changing in time as the cascade develops due to, for example, an adiabatic expansion of the fireball in the source, or as cosmic rays move through cosmic distances with the dilution of the cosmic microwave background. The treatment of other forms of ICTMCs relevant for UHECR energy losses in astrophysical scenarios are discussed more in depth in an upcoming publication.

\section{Example of compound distribution}

\begin{figure}
    \centering
    \includegraphics[width=0.45\linewidth]{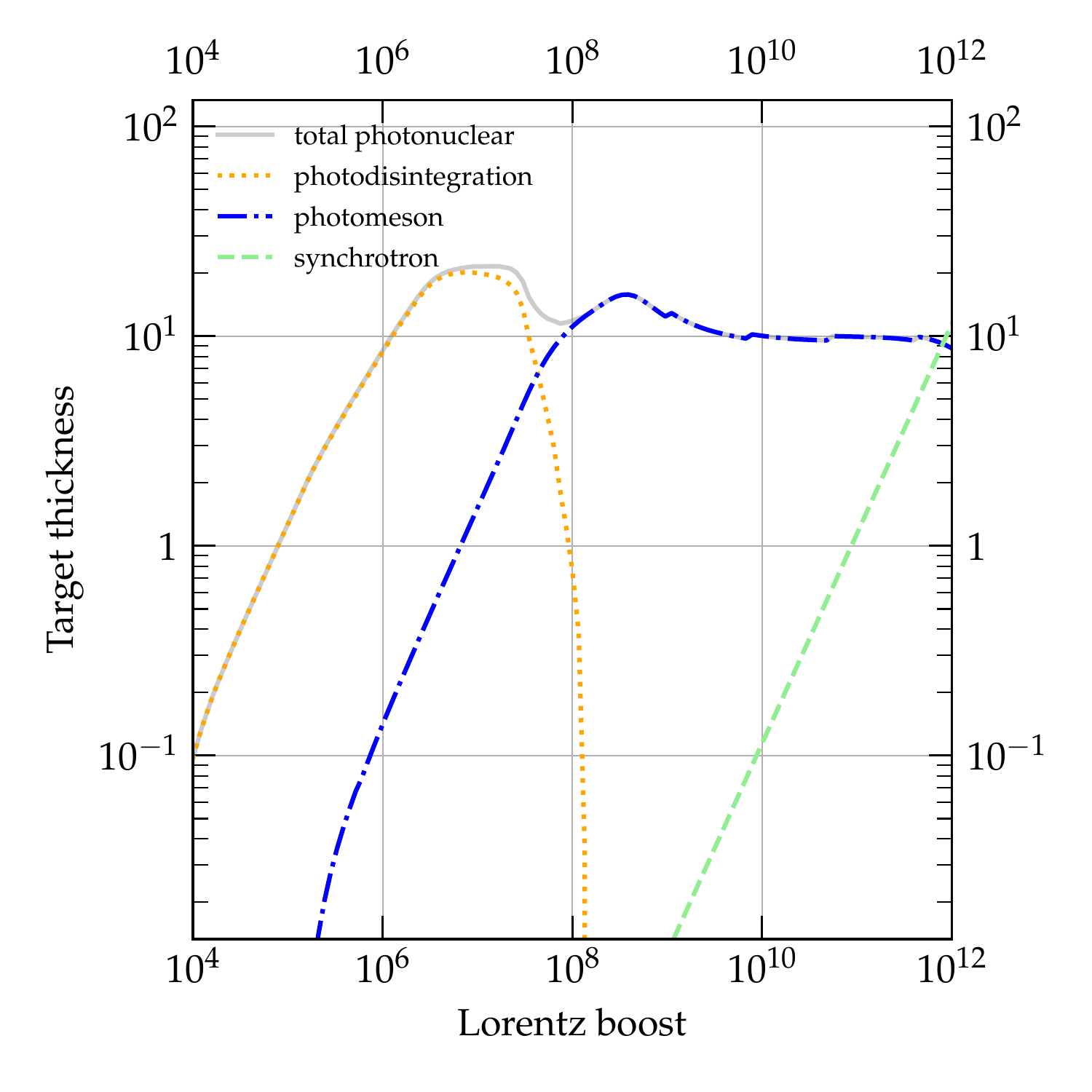}
    \includegraphics[width=0.45\linewidth]{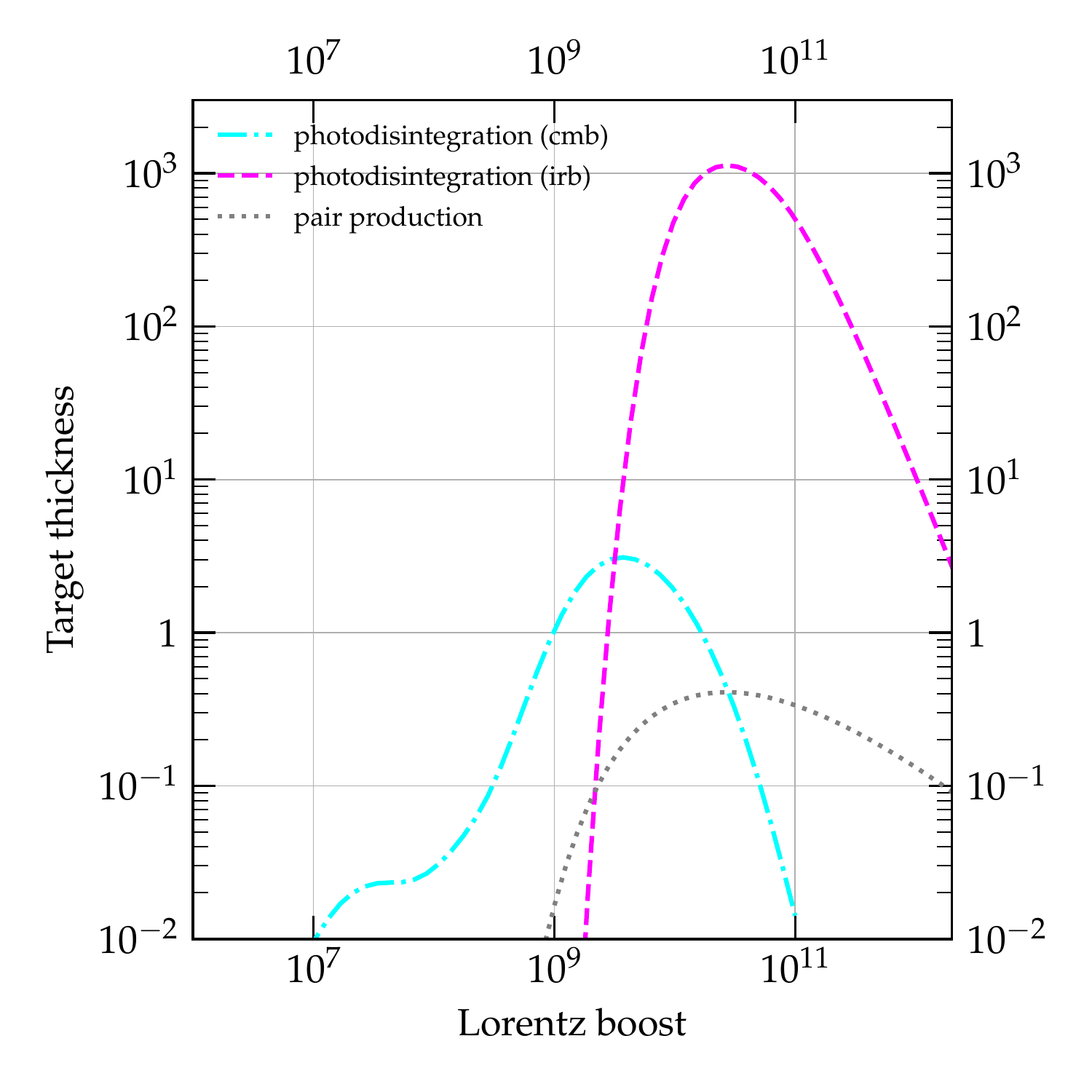}
    \caption{Target thickness in the source (left) and in extragalactic propagation (right) for the injected iron nuclei. The crossed distances assumed are $0.16$~mpc in-source and $40$~Mpc in extragalactic space. Interactions in extragalactic space were separated by target photons: cosmic microwave background (magenta) and infrared background (cyan). Synchrotron and pair production losses are negligible and were ignore for the example shown in Figure~\ref{fig:completepdf}.}
    \label{fig:grbrates}
\end{figure}

To illustrate the compound probability distribution we use, as example of source, a gamma-ray burst located at 40~Mpc from Earth, inspired by the event GRB170817A~\cite{2017ApJ...848L..13A} but modeled as a long gamma-ray burst with parameters chosen conveniently for illustration. The in-source environment is based on the internal shock scenario, as described in \cite{Biehl2018} where the emission region is filled with a homogeneous isotropic photon field such that the luminosity is $10^{51}$ erg/s and the spectral energy distribution is characterized by a broken power-law parametrized as in \cite{Morejon2019}: sub-break spectral index $\alpha_1=-1$, index above the break $\alpha_2=-2$, and 100~eV, 300~keV, 1~keV for the minimum, maximum and break energies respectively. The resulting interaction rates for the injected iron-56 nuclei inside the source are shown in Figure~\ref{fig:grbrates}.

\begin{figure}
    \includegraphics[trim={0 0 2148 0},clip,height=.49\linewidth]{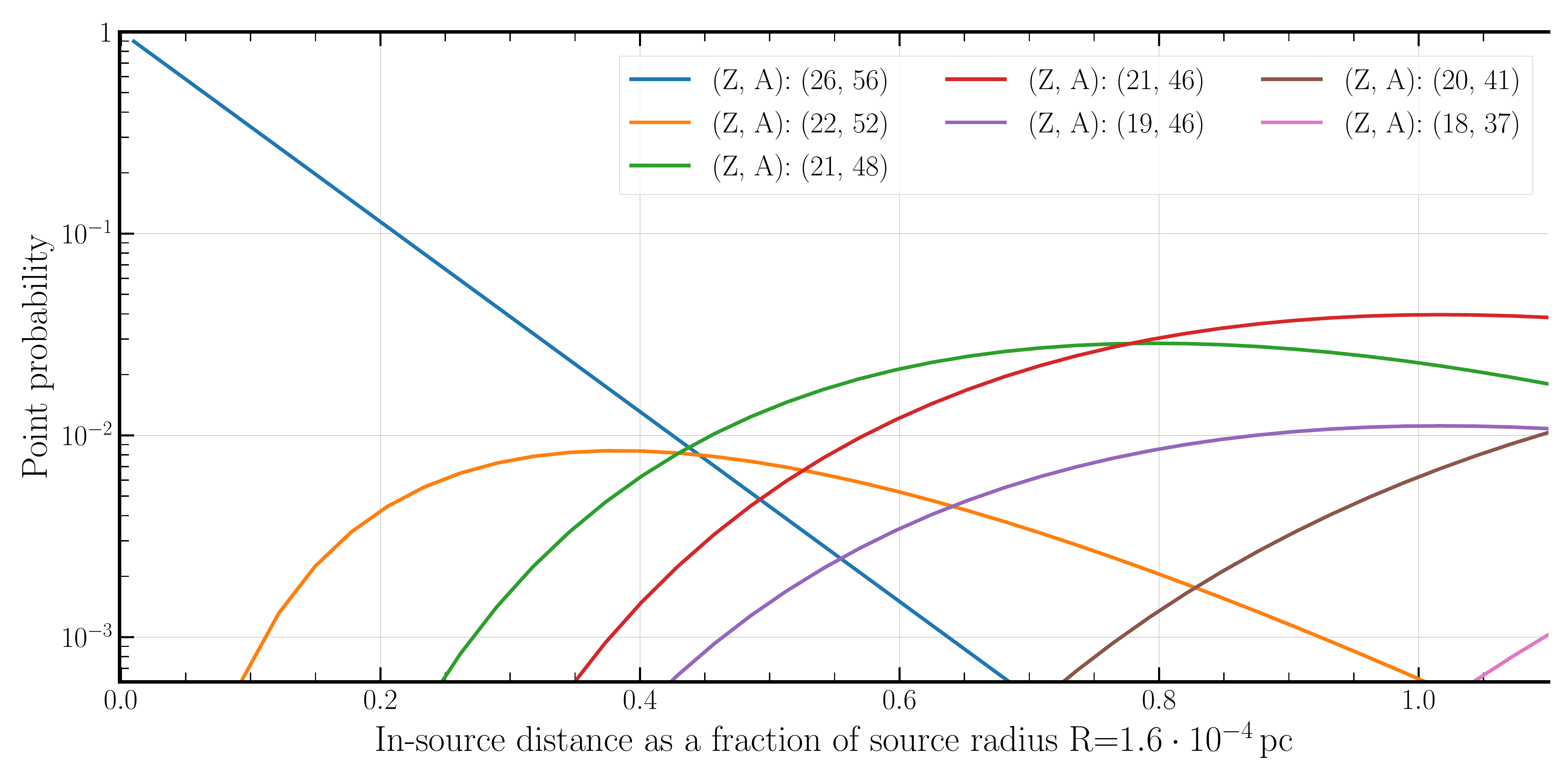}

    \begin{flushright}
        \includegraphics[trim={2148 0 0 0},clip,height=.49\linewidth]{figures/source_to_earth_point_probs_wide.pdf}
    \end{flushright}
    
    \caption{Evolution of point probabilities for some nuclear species (see legend) from source to Earth. The in-source evolution (top) spans distances of fractions of miliparsecs (a few hours at the speed of light), while the extragalactic propagation (bottom) spans tens of megaparsecs (more than three million light-years).}
    \label{fig:completepdf}
\end{figure}

The evolution of the point probabilities for boost $\gamma = 2.4 \cdot 10^9$ is shown in Figure~\ref{fig:completepdf} for the whole path length from the source to the Earth. The top panel shows how the occupation probabilities evolve within the source environment (sub-miliparsec scales), and the bottom panel shows the evolution during extra-galactic propagation (megaparsec scales). The probability distributions are sensitive to such differences in distance scales because the relevant quantity describing the evolution of the system, the traversed thickness, is comparable in magnitude due to the much larger photon density within the source in comparison to the density of the infrared background in the extra-galactic medium (which is the typical target for UHECRs at this boost). The effects of magnetic fields would be reflected in an increase of the propagation path lengths with a corresponding change in the occupation probabilities due to the larger traversed thickness.

\begin{figure}
    \centering
    \includegraphics[width=0.75\linewidth]{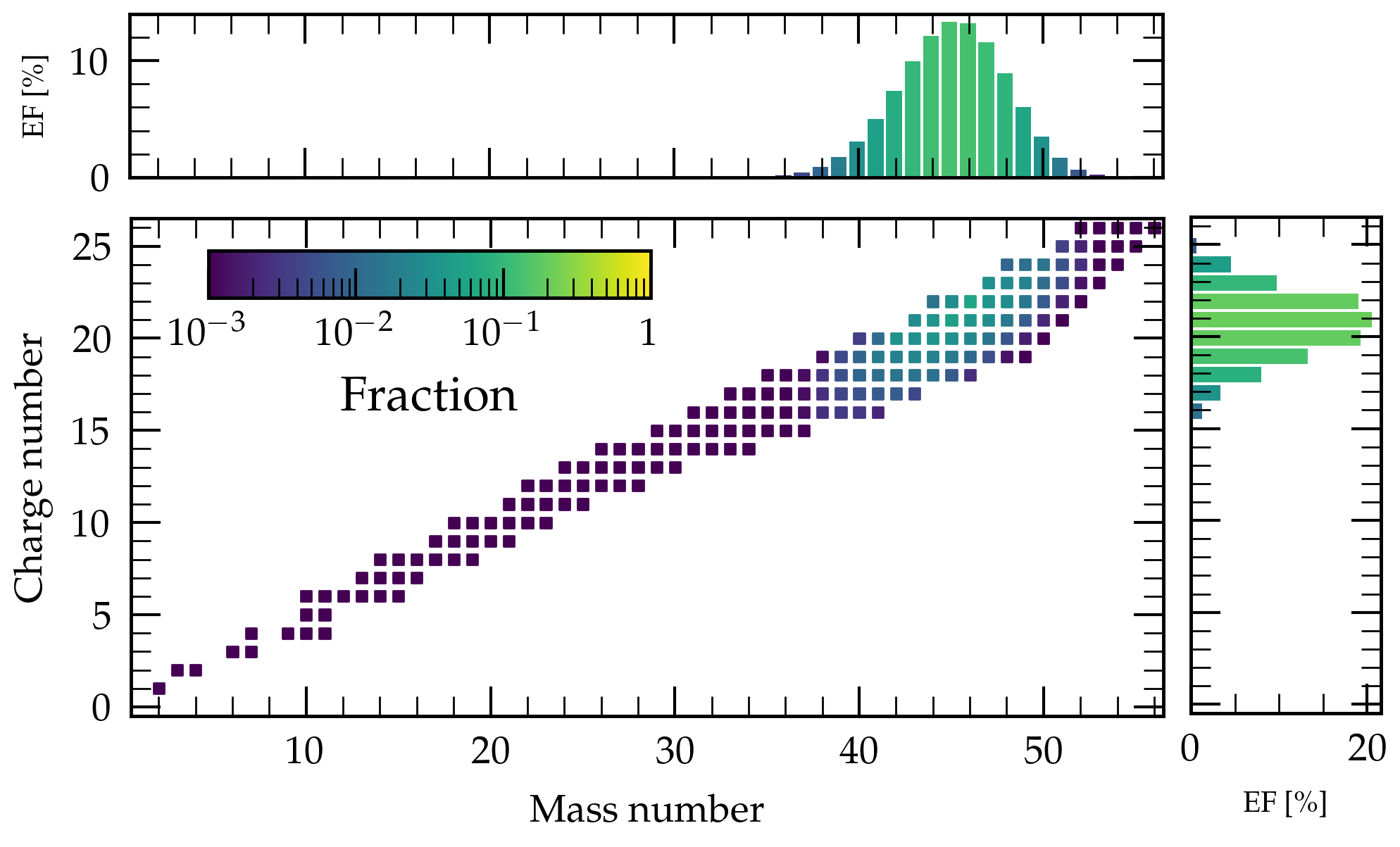}
   \includegraphics[width=0.75\linewidth]{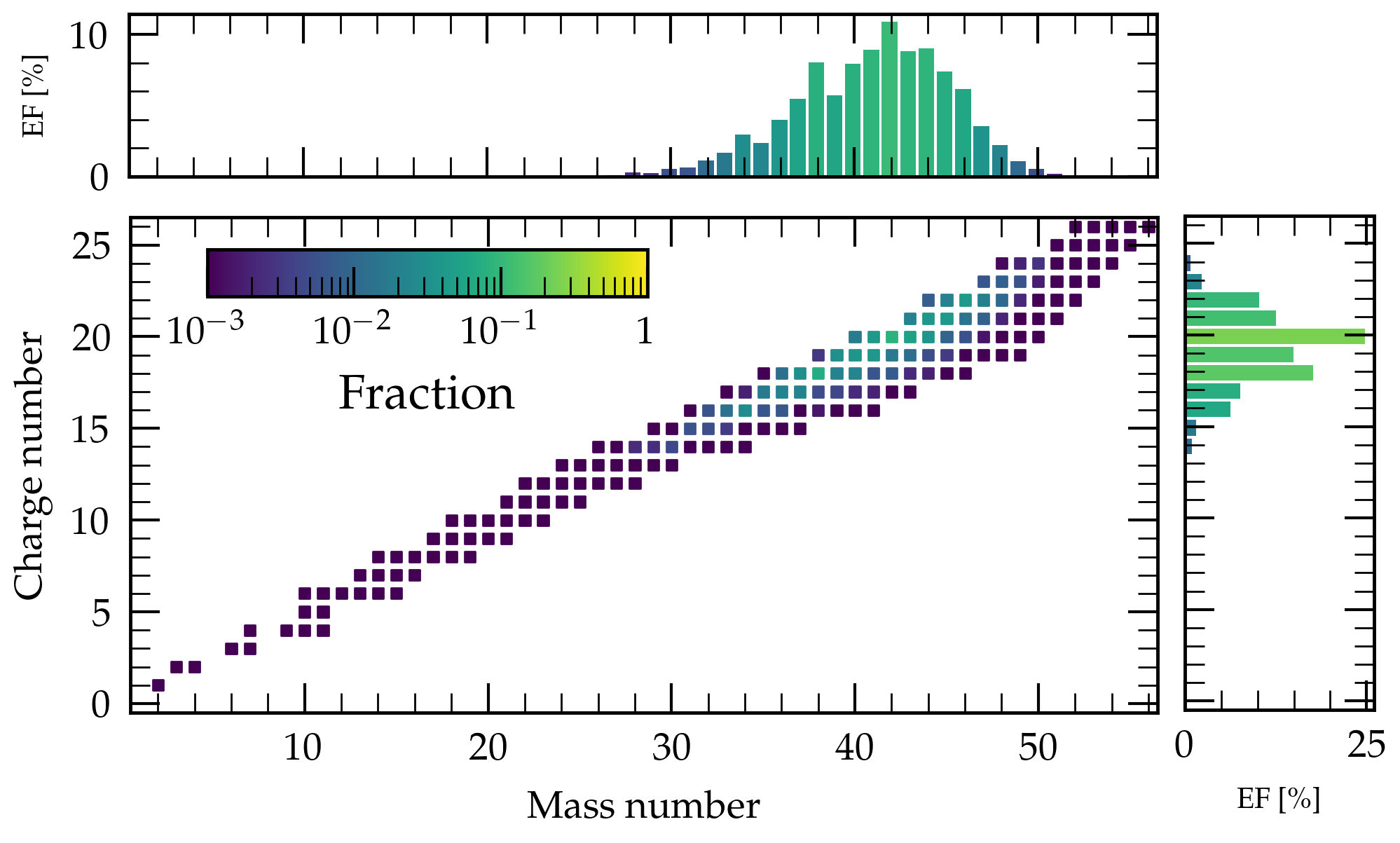}
    \caption{Point probabilities for each nuclear species escaping the source (after in-source propagation, top figure) and upon arrival at Earth (after extra-galactic propagation, bottom). The escaping fractions are given marginally for charge (right) and mass (top) numbers as a percentage of the total.}
    \label{fig:pointprobSourceEarth}
\end{figure}

Figure~\ref{fig:pointprobSourceEarth} presents the occupation probabilities for each of the 184 nuclear species included in the cascade (the cross sections employed in CRPropa~\cite{Batista2016} have been used). The top plot shows the snapshot corresponding to the thickness of the source environment ($0.16$~mpc), while the bottom plot presents the distribution at end of the path, when the cosmic rays arrive at Earth after traveling $40$~Mpc. The modest changes of the occupation probabilities during propagation are due to the lower thickness of the extragalactic medium for the chosen boost.

\section{Potential for locating sources}
Finally, some comments on the application of the stochastic description for locating sources of UHECRs. The reconstruction of the propagation path of a cosmic ray, based on the measured arrival direction and rigidity, can be achieved with codes like CRPropa~\cite{Batista2016} in two ways: a forward propagation approach, or backward propagation approach. The former requires certain assumptions of the composition and energy at the origin but allows taking into account the energy losses and interactions. The latter backtracks the UHECR through the magnetic field and requires no knowledge of the source, although it cannot account for interactions nor the corresponding changes in the path due to rigidity changes (by energy losses or transformations of the nuclear species). Neither of these approaches is well suited for the task of reconstructing the UHECR origin: the latter lacks the important effects of energy losses, and the former requires much computation time (trying different starting conditions) followed by \textit{a posteriori} inference~\cite{Bourriche2024BeyondParticle} for which a broad parameter phase-space is needed for a reliable inference.

\begin{figure}
  \centering
  \begin{tabular}{ c }
    \includegraphics[width=.39\linewidth]{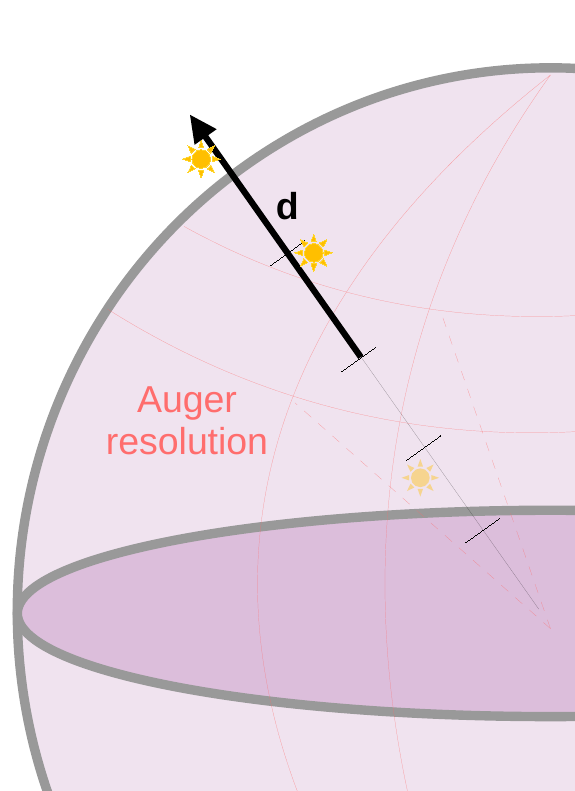}
  \end{tabular}
  \begin{tabular}{ c }
    \includegraphics[width=.53\linewidth]{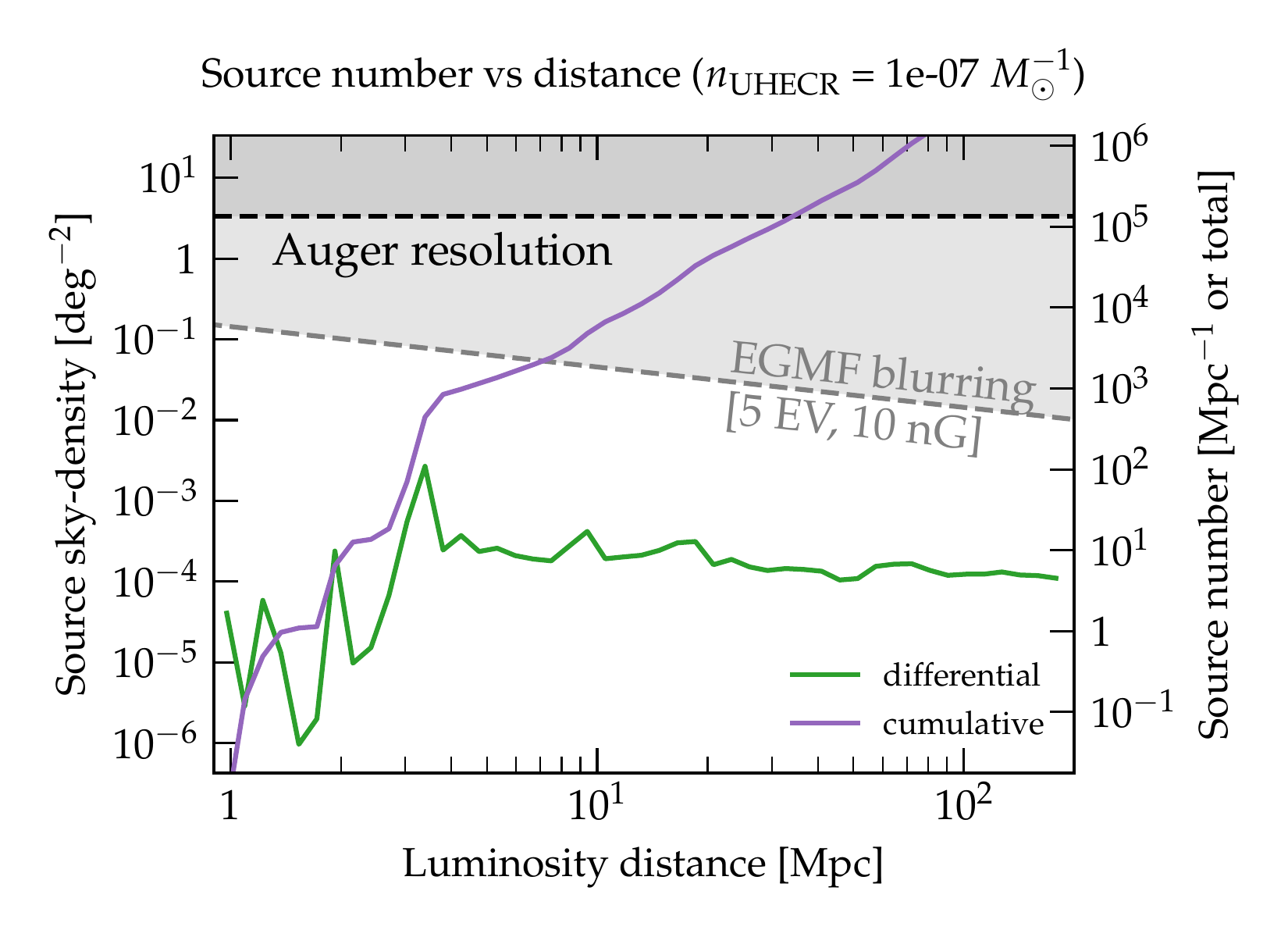} \\
      \includegraphics[width=.53\linewidth]{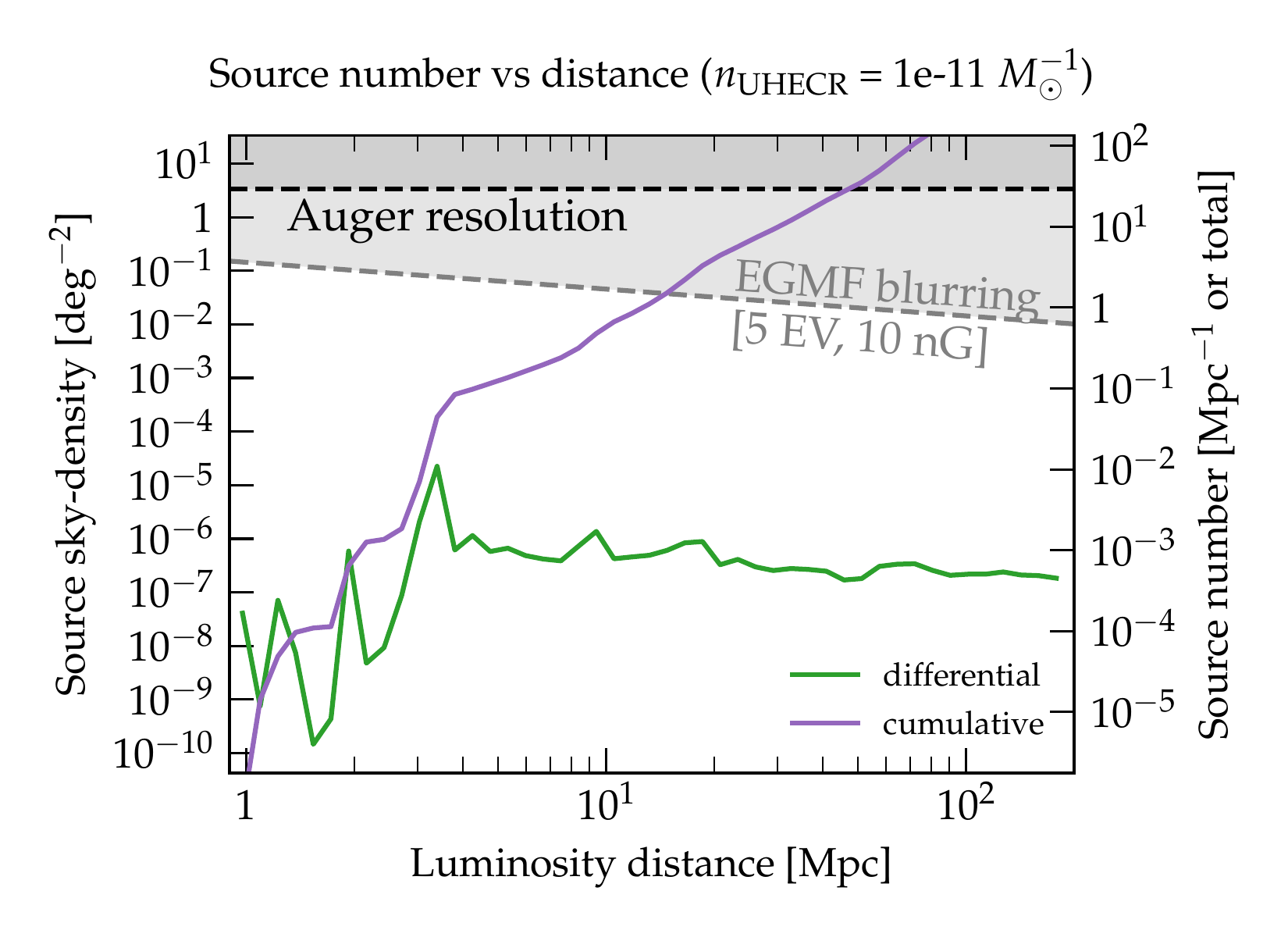} \\
  \end{tabular}
  \caption{Left: schematic representation of the limited source number as a function of the distance and the angular separation in the sky. The resolution of the Pierre Auger Observatory is shown by pairs of parallels in the spherical projection of the sky. Right: average density of UHECR sources in the sky compared to angular resolution of the Pierre Auger Observatory. The fraction of UHECR sources per stellar mass density assumed is $10^{-7} M_{\odot}^{-1}$ (top) and $10^{-11} M_{\odot}^{-1}$ (bottom).}
  \label{fig:visual_sky_density}
\end{figure}

In contrast, the stochastic analytic description is well suited to reconstruct the composition as a function of the path distance, provided that the path is computed suitably (e.g. by using backtracking in CRPropa). Thus, it is a possible future addition to the CRPropa code as a complement for the backtracking approach. In the particular case of extreme-energy cosmic rays (ExECRs), such as the Amaterasu event~\citep{doi:10.1126/science.abo5095}, the propagation paths can be better constrained due to interactions with the CMB, and the localization of the origin can be as narrow as a few to tens of megaparsecs \citep{Unger2024WhereFrom,Bourriche2024BeyondParticle,Morejon:2024rtq} depending on the composition. Furthermore, the original composition is not needed for a good reconstruction of the disintegration cascade as the intermediate mass groups appear at similar distances~\cite{Morejon:2024rtq}, thus the observed composition is sufficient to narrow the distance to the origin. The resolution power for the origin can further be estimated with a back-of-the-envelope calculation for the angular resolution: assuming that the UHECR source density is proportional to the stellar mass density, the number of sources in the sky within certain a distance from the Milky Way can be computed. Figure~\ref{fig:visual_sky_density} (right) illustrates such estimates based on the catalog of \cite{Biteau2021} to estimate the stellar mass volumetric density $M_*$ as a function of luminosity distance and different values of the fraction $n_{\mathrm{UHECR}}$ of UHECR sources per stellar mass. The mean number of sources in the sky $\mu(d)$ limited to a luminosity distance $d$ is estimated as $\mu(d)=n_{\mathrm{UHECR}} \int_{1\,{\mathrm{Mpc}}}^d r^2 M_*(r) dr$ and represented with a purple line in Figure~\ref{fig:visual_sky_density} (the differential is shown by the green line). With the best resolution of the Pierre Auger Observatory ( $0.3\deg$~\cite{2009NuPhS.190...20B}) a minimal sky-resolution of $0.33\deg^2$ is obtained (dashed black line) which represents a limit on our experimental capability to separate sources in the sky, i.e. values of sky density lower than the limited are favorable for angular separation of sources. A possibly more stringent limit results from estimating the deflections on the extra-galactic magnetic field (EGMF) $\Delta \theta \approx \frac{2}{\pi}\frac{B}{R}\sqrt{\lambda_Bd}$~~\cite{Lee1995}, however the strength of the EGMF is not well constrained. This limit is shown (gray dashed line) for cosmic rays with $R=5$~EV rigidity and a $B=10$~nG intensity EGMF field with a coherence length of $\lambda_B=10$~kpc.
These limits suggest the most promising volume for the search of UHECR sources: the radius where the sources can be resolved (distance where the sky-density matches the cumulative source number) can be probed with ExECRs of sufficiently large rigidity, whose path distance can be resolved in composition by the stochastic analytic description~\cite{Morejon:2023cpe,Morejon:2024rtq}. The additional effect on the composition due to propagation in the source environment can also be included in the stochastic analytic description as discussed in the previous section.

\section*{Acknowledgements}
The authors would like to thank Jannis Pawlowsky for fruitful discussions and for providing the schematic representation of the sky density in Figure~\ref{fig:visual_sky_density} (left). This work has received funding via the grant MultI-messenger probe of Cosmic Ray Origins (MICRO) from the DFG through project number 445990517.

\bibliographystyle{apalike}
\setlength{\bibsep}{0pt plus 0.3ex}
\bibliography{references}

\end{document}